\documentclass[twocolumn,showpacs,prb,amsmath,amssymb,floatfix]{revtex4-1}
\usepackage{amsmath,amssymb,color}
\usepackage{bm}
\usepackage{epsfig}
\usepackage{psfrag}

\newcommand{\bd}{\bm}

\def\Xint#1{\mathchoice
   {\XXint\displaystyle\textstyle{#1}}%
   {\XXint\textstyle\scriptstyle{#1}}%
   {\XXint\scriptstyle\scriptscriptstyle{#1}}%
   {\XXint\scriptscriptstyle\scriptscriptstyle{#1}}%
   \!\int}
\def\XXint#1#2#3{{\setbox0=\hbox{$#1{#2#3}{\int}$}
     \vcenter{\hbox{$#2#3$}}\kern-.5\wd0}}

\def\dashint{\Xint-}

\begin{document}

\title{Elastic constants and ultrasound attenuation in the spin-liquid phase of Cs$_2$CuCl$_4$  
}

\author{S. Streib}
\affiliation{Institut f\"{u}r Theoretische Physik, Universit\"{a}t Frankfurt,  Max-von-Laue Strasse 1, 60438 Frankfurt, Germany}

\author{P. Kopietz}
\affiliation{Institut f\"{u}r Theoretische Physik, Universit\"{a}t
  Frankfurt,  Max-von-Laue Strasse 1, 60438 Frankfurt, Germany}

\author{P. T. Cong}
\affiliation{Physikalisches Institut, Universit\"{a}t
  Frankfurt,  Max-von-Laue Strasse 1, 60438 Frankfurt, Germany}

\author{B. Wolf}
\affiliation{Physikalisches Institut, Universit\"{a}t
  Frankfurt,  Max-von-Laue Strasse 1, 60438 Frankfurt, Germany}

\author{M. Lang}
\affiliation{Physikalisches Institut, Universit\"{a}t
  Frankfurt,  Max-von-Laue Strasse 1, 60438 Frankfurt, Germany}

\author{N. van Well}
\affiliation{Physikalisches Institut, Universit\"{a}t
  Frankfurt,  Max-von-Laue Strasse 1, 60438 Frankfurt, Germany}

\author{F. Ritter}
\affiliation{Physikalisches Institut, Universit\"{a}t
  Frankfurt,  Max-von-Laue Strasse 1, 60438 Frankfurt, Germany}

\author{W. Assmus}
\affiliation{Physikalisches Institut, Universit\"{a}t
  Frankfurt,  Max-von-Laue Strasse 1, 60438 Frankfurt, Germany}

\date{January 16, 2015}

\begin{abstract}

The spin excitations
in the spin-liquid phase of the anisotropic
triangular lattice quantum antiferromagnet Cs$_2$CuCl$_4$
have been shown to propagate dominantly along the
crystallographic $b$ axis.
To test this dimensional reduction scenario,
we have performed ultrasound experiments
in the spin-liquid phase of Cs$_2$CuCl$_4$
probing the elastic constant $c_{22}$ and the sound attenuation
along the $b$ axis as a function of an external magnetic field along the $a$ axis.
We show that our data can be
quantitatively explained within the framework
of a nearest-neighbor spin-$1/2$ Heisenberg chain, where fermions are introduced via the Jordan-Wigner transformation and
the spin-phonon interaction arises from the usual exchange-striction
mechanism.

\end{abstract}


\maketitle

\begin{figure}[t]
\includegraphics[width=80mm]{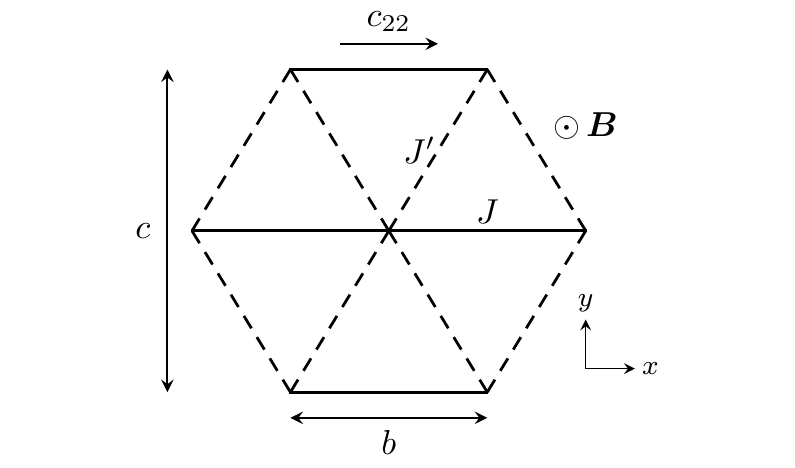}
  \caption{%
Part of the  anisotropic triangular lattice formed by the
spins in  Cs$_2$CuCl$_4$. The largest exchange coupling $J$
connects nearest-neighbor spins along the
crystallographic $b$ axis. The corresponding
elastic constant is denoted by $c_{22}$.
In this work we consider
only the case where the magnetic field $\bd{B}$ is along the $a$ axis perpendicular to the
plane of the lattice.
}
  \label{fig:lattice}
\end{figure}

%\section{Introduction}
Spin-liquid behavior can occur either in a spin-liquid ground state or in a spin-liquid phase at finite temperatures.
One of the characteristic properties of spin liquids are strong short-range spin correlations in the absence of long-range magnetic order.
Such a behavior has been observed in
the magnetic insulator Cs$_2$CuCl$_4$, for example in inelastic neutron scattering experiments,\cite{Coldea03}
at temperatures between $0.6\;  {\rm K}$ and $2.6\;  {\rm K}$  in magnetic fields below the saturation field $B_c = 8.5 \, {\rm T}$.
Experimentally, the boundary between the  spin-liquid phase and the conventional paramagnetic phase has been characterized by broad peaks in the specific heat \cite{Radu05} and in the magnetic susceptibility. \cite{Tokiwa06} 
Cs$_2$CuCl$_4$ can be modeled by a spin-$1/2$
Heisenberg antiferromagnet on a spatially  anisotropic triangular lattice
with nearest-neighbor exchange couplings $ J = 4.34\;  {\rm K} $ 
along the crystallographic $b$ axis
and $J^{\prime} = 1.49 \; {\rm K} \approx J/3 $ along the diagonal links within the $bc$ plane (see Fig.~\ref{fig:lattice}).\cite{Coldea02} The interplane interaction $J''=0.2\;  {\rm K}$ and the Dzyaloshinskii-Moriya interaction $D=0.23\;  {\rm K}$ can be neglected in the temperature range of the spin-liquid phase.
Given the fact that the diagonal coupling $J'$ is nonnegligible, one would naively expect 
an anisotropic  two-dimensional spin liquid state. 
However, several independent calculations \cite{Weng06,Yunoki06,Hayashi07,Starykh07,Kohno07,Kohno09,Heidarian09,Tay10,
Starykh10,Herfurth13, Tocchio14} found that the spin excitations in the anisotropic triangular lattice antiferromagnet are quasi-one-dimensional and propagate dominantly along the direction corresponding to the largest
exchange coupling, which is the crystallographic $b$ axis  in Cs$_2$CuCl$_4$. 
In this work we shall give further evidence for this dimensional reduction
scenario \cite{Balents10} by showing that ultrasound experiments probing the
sound propagation along the $b$ axis can be quantitatively explained
using a one-dimensional Heisenberg chain which is coupled to lattice vibrations
via the usual exchange-striction mechanism.\cite{Luethi05}

The spin-phonon interaction and the ultrasonic attenuation in two-dimensional 
spin liquids have recently been discussed
by Zhou and Lee,\cite{Zhou11} and by Serbyn and Lee.\cite{Serbyn13}
In one-dimensional Heisenberg \cite{Pytte74}  and $XY$ chains \cite{Lima83} 
the interaction between the spin degrees of freedom and the 
phonons
were studied a long time ago, but
these older works mainly focused on the spin-Peierls
transition and treated the spin-phonon interaction in an adiabatic
approximation. 
Moreover, only the first derivative of the exchange coupling with respect to the phonon coordinates were considered, which turns out to be insufficient to explain our ultrasound experiments 
for the $c_{22}$ mode in Cs$_2$CuCl$_4$.
 
From the exact Bethe ansatz solution of the
spin-$1/2$ antiferromagnetic Heisenberg chain we know that the ground state is
a spin liquid, exhibiting
algebraic correlations but no long-range magnetic order.
The elementary excitations above this ground state are spinons carrying spin $1/2$. A combination of numerical and analytical
methods has lead to an excellent understanding of this model;\cite{Giamarchi03} for example,
exact numerical results for the magnetization,\cite{Griffiths64} magnetic susceptibility, \cite{Eggert94} specific heat,\cite{Kluemper98} and the dynamic structure factor \cite{Mourigal13,Lake13} are available. However, a proper microscopic calculation of ultrasound propagation and attenuation in the Heisenberg chain cannot be found
in the literature. Below we shall present a simple
solution of this problem using the Jordan-Wigner representation
of the spin-algebra in terms of spinless fermions.\cite{Lieb61}
We shall also present some new data
of the elastic constant $c_{22}$ and the corresponding ultrasound damping rate
in the spin-liquid phase of
Cs$_2$CuCl$_4$ which agree very well with our theory. For details concerning the experiment and the sample preparation we refer to Ref.~[\onlinecite{Kreisel11}] and Ref.~[\onlinecite{Krueger10}], respectively. The ultrasound physics of Cs$_2$CuCl$_4$ has  previously been studied for magnetic fields along the $a$ axis in the ordered phase using spin-wave theory, \cite{Kreisel11} and for magnetic fields along the $b$ axis in the spin-liquid phase by combining phenomenological expressions for the ultrasound propagation and attenuation with calculations for two-dimensional spin models.\cite{Sytcheva09}

Assuming that the relevant spin excitations can propagate only
along the crystallographic $b$ axis in the spin-liquid phase of Cs$_2$CuCl$_4$,
we expect that ultrasound experiments probing the $c_{22}$ mode along the $b$ axis can be explained by the following one-dimensional spin-phonon Hamiltonian,
\begin{eqnarray}
{\cal{H}} & = & \sum_n  J_n 
 \left[ {\bd{S}}_n \cdot {\bd{S}}_{n+1} - 1/4 \right]   - h \sum_n 
 S^z_n + {\cal{H}}_2^p,
 \label{eq:hamiltonian}
 \end{eqnarray} 
where $\bd{S}_n$ are spin-$1/2$ operators localized at the positions
$x_n$ on a chain with $N$ spins and periodic boundary conditions, and 
$h=g\mu_B B$ (with $g\approx 2.2$)\cite{Coldea02} is the Zeeman energy associated with an external magnetic field $\bd{B}$
along the crystallographic $a$ axis; see Fig.~\ref{fig:lattice}.
The spin-phonon coupling arises from the fact that for a
vibrating lattice the spins are located at
$x_n = n b + X_n$, where $nb$ (with $n = 1, \ldots , N$) are the
points of a one-dimensional lattice with lattice spacing $b$,
and $X_n$ denote the deviations from
the lattice points. Since the exchange coupling
$J_n$ between a pair of spins $\bd{S}_n$ and $\bd{S}_{n+1}$
located at $x_n$ and $x_{n+1}$ depends on the actual  distance
between the spins, $J_n$ is a function of $X_{n+1} - X_n$.
Assuming that this difference is small, we may expand to second order,
 \begin{equation}
 J_n  \approx  J + J^{(1)} ( X_{n+1} - X_n ) + 
\frac{ J^{(2)}}{2}
 ( X_{n+1} - X_n )^2,
 \label{eq:Jexp}
 \end{equation}
where $J^{(1)}$ and $J^{(2)}$ are the first and second derivative 
of the exchange coupling along the $b$ axis with respect to the
phonon coordinates.
As usual, we quantize the lattice vibrations by introducing
the conjugate momenta  $P_n$ and demanding that
$[X_n, P_m ] = i \delta_{n,m}$, where we have set $\hbar =1$.
The last term in Eq.~(\ref{eq:hamiltonian})
describes non-interacting phonons with dispersion $\omega_q = c | q |/b$, \cite{footnote1}
 \begin{equation}
 {\cal{H}}_2^p =
 \sum_{q} \left[ \frac{ P_{-q} P_q }{2M} +  \frac{M}{2} \omega_q^2 X_{-q } X_q
  \right],
 \end{equation}
where $M$ is the mass attached to the vibrating sites, and
the operators $X_q$ and $P_q$ are defined via 
the  Fourier expansions $X_n = N^{-1/2} \sum_q e^{ i q n} X_q$ and
$P_n = N^{-1/2} \sum_q e^{ i q n } P_q$; the phonon momentum $q$ is given in units of the inverse lattice spacing $1/b$.

\begin{figure}[t]
\includegraphics[width=80mm]{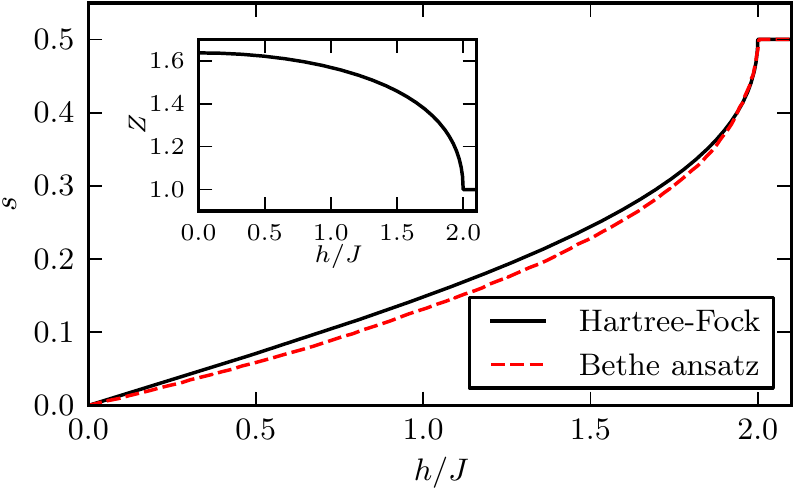}
  \caption{%
(Color online)
Comparison of the Hartree-Fock result for the 
magnetization curve $s (h)$ of the Heisenberg chain (without phonons) at $T=0$ with the exact Bethe ansatz.\cite{Griffiths64} Inset: Renormalization $Z(h)$ of the nearest-neighbor hopping.
}
\label{fig:magnetization}
\end{figure}

To explain ultrasound experiments, we calculate the 
self-energy correction $\Pi ( q , i \omega )$ to the phonon 
propagator, which arises from the coupling
of the phonons to the spins.
Let us therefore represent the spin operators in terms of spinless
fermions  using the Jordan-Wigner transformation \cite{Lieb61}
% \begin{subequations}
 \begin{equation}
 S^+_n  = (S^-_n)^{\dagger} = c^{\dagger}_n (-1)^n e^{i \pi \sum_{ j < n } c^{\dagger}_j c_j }, \; \; \; 
% \\
 S^z_n  =  c^{\dagger}_n c_n - 1/2,
 \end{equation}
% \end{subequations}
where $c_n$ annihilates a fermion at site $x_n$ and
the phase factor $(-1)^n$ is introduced for convenience.
Our spin-phonon Hamiltonian (\ref{eq:hamiltonian}) then assumes the form
 \begin{eqnarray}
 {\cal{H}}
 & = & - \frac{1}{2}  \sum_n J_n (  c^{\dagger}_n c_{n+1} + c^{\dagger}_{n+1} c_n   
  +  c^{\dagger}_n c_{n} + c^{\dagger}_{n+1} c_{n+1} )  
 \nonumber
 \\
 &+ &  \sum_n J_n c^{\dagger}_n c_n c^{\dagger}_{n+1} c_{n+1}
 - h \sum_n c^{\dagger}_n c_n   + N h/2 + {\cal{H}}_2^p.
 \hspace{7mm}
 \label{eq:hamiltonianJW}
 \end{eqnarray}
In this work, we shall treat the two-body interaction in the second line of 
Eq.~(\ref{eq:hamiltonianJW}) within the self-consistent Hartree-Fock approximation, which amounts to approximating the two-body term  by
 \begin{eqnarray}
 c^{\dagger}_n c_n c^{\dagger}_{n+1} c_{n+1}
 & \approx &  
\rho  ( c^{\dagger}_{n+1} c_{n+1} +   c^{\dagger}_n c_n)  - \rho^2 
 \nonumber
 \\
 &- &   \tau ( c^{\dagger}_n c_{n+1} + 
 c^{\dagger}_{n+1} c_{n} ) + \tau^2,
 \label{eq:HF}
 \end{eqnarray}
where the dimensionless variational parameters $\rho$ and $\tau$ satisfy
the self-consistency conditions
 \begin{equation}
 \rho = \langle c^{\dagger}_n c_n \rangle, \; \; \;
 \tau = \langle c^{\dagger}_n c_{n+1} \rangle.
 \end{equation}
In the absence of phonons, the solution of these equations 
was worked out a long time ago by Bulaevskii.\cite{Bulaevskii62}
Within the Hartree-Fock
approximation the fermion dispersion is 
% \begin{equation}
 $\xi_k = - Z J  \cos k + 2 s  J  - h$,
% \end{equation}
where $Z = 1 + 2 \tau$ is the dimensionless
renormalization factor of the nearest-neighbor hopping,  
$s = \rho -1/2$ is the dimensionless magnetization,
and $k$ is the fermion lattice momentum in units of the inverse lattice spacing
$1/b$.
In Fig.~\ref{fig:magnetization} we show the numerical result for $Z(h)$ at $T=0$ and we compare our mean-field result for $s(h)$ with the exact
magnetization curve of the Heisenberg chain obtained via the 
Bethe ansatz.\cite{Griffiths64}

\begin{figure}[t]
\includegraphics[width=80mm]{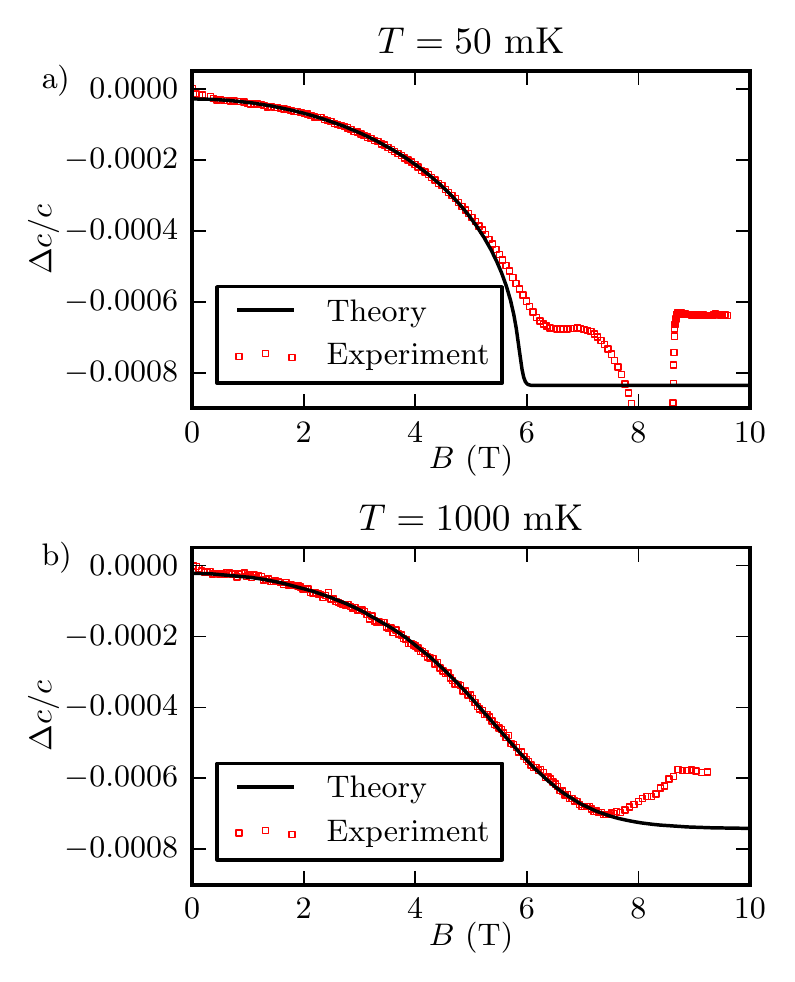}
  \caption{%
(Color online) Comparison of theory and experiment for the relative change of the sound velocity of the $c_{22}$ mode: a) in the ordered phase ($T=50\,{\rm mK}$) with coupling constants $g_1=0$ and $g_2=-1.2\times10^{-3}$, b) in the spin-liquid phase ($T=1\,{\rm K}$) with coupling constants $g_1=0$ and $g_2=-1.1\times10^{-3}$. In the fitting procedure we have allowed a constant offset for $\Delta c/c$ which is necessary due to a small anomaly in the experimental data close to zero magnetic field.
}
  \label{fig:fitc}
\end{figure}

To obtain the change of the elastic constant and  the sound attenuation,
we should calculate the self-energy of the phonons due to the coupling to the 
spins.
Substituting the gradient expansion (\ref{eq:Jexp}) 
for the exchange coupling and the 
Hartree-Fock decoupling (\ref{eq:HF}) into
Eq.~(\ref{eq:hamiltonianJW}), 
we arrive at the approximate 
spin-phonon Hamiltonian
 \begin{equation}
 {\cal{H}}  =  F_0 + \sum_k \xi_k c^{\dagger}_k c_k  + {\cal{H}}_2^p
 + \delta {\cal{H}}_2^p + {\cal{H}}_3^{sp} + {\cal{H}}_4^{ sp},
 \end{equation}
where $F_0 / N  = h/2 + J ( \tau^2 - \rho^2 )$ and 
\begin{subequations}
 \begin{eqnarray}
 \delta {\cal{H}}_2^p & = & 2 J^{(2)}  ( \tau^2 - \rho^2)
 \sum_q  \sin^2 (q/2) X_{-q} X_q ,
 \\
 {\cal{H}}_3^{sp} & = & \frac{1}{\sqrt{N}} \sum_{ k^{\prime} k q }
 \delta^{\ast}_{ k^{\prime} , k+q } \Gamma_3 (  k , q ) c^{\dagger}_{ k^{\prime}}
 c_{k} X_q ,
 \label{eq:H3sp}
 \\
 {\cal{H}}_4^{sp} & = & \frac{1}{2 {N}} \sum_{ k^{\prime} k q_1 q_2 }
 \delta^{\ast}_{ k^{\prime} , k+q_1 + q_2 } \Gamma_4 ( k , q_1 , q_2  ) 
 c^{\dagger}_{ k^{\prime}} c_{k} X_{q_1} X_{q_2 } .
 \nonumber
 \\
 & &
 \label{eq:H4sp}
 \end{eqnarray}
 \end{subequations}
Here  $\delta^{\ast}_{ k^{\prime} , k } = \sum_m \delta_{ k^{\prime} , k + 2 \pi m}$
enforces momentum conservation modulo a vector of the reciprocal lattice,  
$c_k  =  N^{-1/2} \sum_n e^{ - i k n } c_n$,
and the cubic and quartic interaction vertices are for small phonon momenta given by
 \begin{subequations}
 \begin{eqnarray}
  \Gamma_3 (  k , q )   
 &  \approx & 
 - i q J^{(1)}  [ Z \cos k 
  -  2 s ],
 \label{eq:Gamma3long}
 \\
  \Gamma_4 (  k , q_1 , q_2  )  
 &  \approx   &  q_1 q_2 J^{(2)}
 [ Z\cos k
 - 2  s ].
 \hspace{7mm}
 \end{eqnarray}
\end{subequations}
The coupling to the fermions gives rise to a momentum- and frequency dependent self-energy correction
$\Pi ( q , i \omega )$ to the propagator of the
phonon field $X_q$, which  is proportional to $[ \omega^2 + \omega_q^2 + \Pi ( q , i \omega )]^{-1}$.
To second order in the derivatives of the exchange coupling
the phonon self-energy has three contributions,
 $
 \Pi ( q , i \omega )  = \Pi_2 ( q ) + \Pi_3 ( q , i \omega ) + \Pi_4 ( q ),
 $
where
 \begin{subequations}
 \begin{eqnarray}
 \Pi_2 ( q ) & = &   [ J^{(2)} / M ] [ \tau^2- \rho^2 ] 
  4 \sin^2 \left( q/2 \right),
 \\
 \Pi_3 ( q , i \omega )  
& = & \frac{1}{MN} \sum_k 
 \frac{ f_k - f_{ k+q}}{ \xi_k - \xi_{k+q} + i \omega}
  | \Gamma_3 ( k, q ) |^2,
 \hspace{7mm}
 \label{eq:Pi3}
 \\
\Pi_4 ( q ) & = & \frac{1}{M  N} \sum_{ k} f_k \Gamma_4 ( k , q,-q),
 \end{eqnarray}
\end{subequations}
and $f_k = [ e^{ \beta \xi_k} +1 ]^{-1}$ is the occupation of the fermion state
with momentum $k$ in self-consistent Hartree-Fock approximation.
From the analytic continuation of the self-energy $\Pi (q , i \omega )$
to real frequencies we obtain  the renormalized phonon energy and the 
phonon damping,\cite{Kreisel11}
 \begin{eqnarray}
 \tilde{\omega}_q & = & \omega_q + \frac{ {\rm Re} \Pi ( q , \omega_q + i 0 )}{2 \omega_q },
 \gamma_q  =  - \frac{ {\rm Im} \Pi ( q , \omega_q + i 0)}{ 2 \omega_q }.
 \hspace{7mm}
 \label{eq:omegagamma}
 \end{eqnarray}
The renormalized phonon velocity can  be obtained from $\tilde{c} /c  = \lim_{ q \rightarrow 0} \tilde{\omega}_q / \omega_q$, which yields for the shift $\Delta c = \tilde{c} - c$,
 \begin{subequations}\label{eq:deltac}
 \begin{eqnarray}
\Delta c/c   & = &  g_1 c^{(1)}+g_2 c^{(2)}, \\ 
c^{(1)}&=&\dashint_{ - \pi}^{\pi} \frac{ dk}{2 \pi}  J  f^{\prime} ( \xi_k ) 
  \frac{   {v}_k  }{ {v}_k - c }
  [ 2s - Z \cos k   ]^2,
 \\
c^{(2)} & = &  s^2 - Z^2/4,
 \end{eqnarray}
\end{subequations}
 where $\dashint$ denotes the Cauchy principal value,
$f^{\prime} ( \xi_k) = - \beta f_k [ 1 -
 f_k]$ 
is the derivative of the Fermi function, 
 $
 {v}_k = Z J b   \sin k$
is the group velocity of the fermionic excitations, and we have introduced
the dimensionless coupling constants $g_1 = [ J^{(1)} b ]^2/(2  M c^2 J )$ and
$g_2 = J^{(2)} b^2 / (2 M c^2 )$.  In principle it should be possible to calculate
these coupling using {\it ab initio} methods, but here we simply determine
$g_1$ and $g_2$ by fitting
our theoretical prediction (\ref{eq:deltac}) to our experimental data.
In Fig.~\ref{fig:fitc} we show a comparison between theory and experiment 
as a function of the magnetic field. 
We find that the relative change of the sound velocity $\Delta c/c$ is best fitted by $g_1\approx0$ and $g_2\approx-1.1\times10^{-3}$, which gives $J^{(2)}b^2\approx-238J$. Only for our experimental data at the highest temperature $T=1.15\,{\rm K}$, we get a finite value $g_1=0.85\times 10^{-3}$, which gives $J^{(1)}b\approx \pm 14J$. Because of the large value of $c/(Jb)\approx6.8$, the term $c^{(1)}$ is more than one order of magnitude smaller than $c^{(2)}$ for Cs$_2$CuCl$_4$.
While in the ordered phase ($T = 50\, {\rm mK}$, upper panel) 
our theoretical prediction (\ref{eq:deltac}) agrees only for magnetic fields up to $5\, {\rm T}$ with our experimental data,
in the spin-liquid phase ($ T = 1 \, {\rm K}$, lower panel) we obtain excellent agreement
between theory and experiment for magnetic fields up to $7 \, {\rm T}$.
A natural explanation for the deviations at larger fields is that
in this regime the fluctuations are controlled by the dilute Bose gas quantum critical 
point at $B_c = 8.5 \, {\rm T}$,\cite{Radu05} which of course cannot be described by our one-dimensional model.

\begin{figure}[t]
\includegraphics[width=80mm]{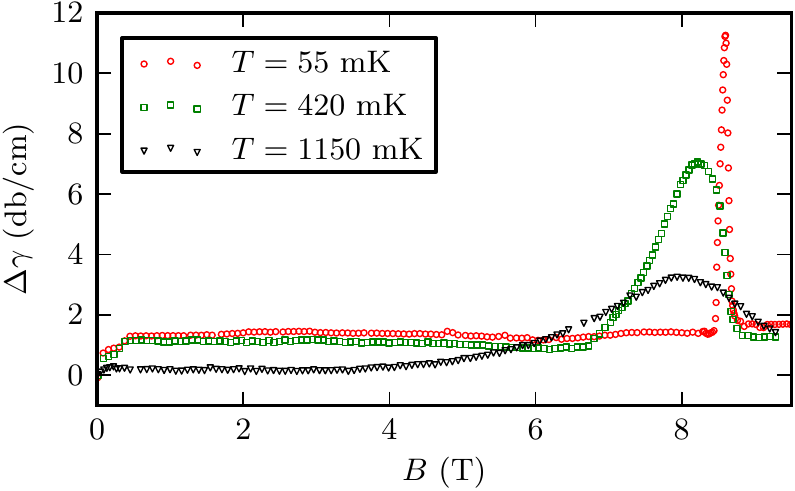}
  \caption{%
(Color online)
Experimental results for the relative change $\Delta \gamma$ of the sound attenuation of the $c_{22}$ mode as a function of the
magnetic field for three different temperatures.
}
  \label{fig:dampexp}
\end{figure}

Finally, let us discuss the ultrasound attenuation 
of the $c_{22}$ mode in  Cs$_2$CuCl$_4$.
Our experimental data for three different temperatures as a function of the magnetic field
are shown in Fig.~\ref{fig:dampexp}.
In the regime $B \lesssim 7 \, {\rm T}$ where  the fluctuations controlled by the quantum critical point are negligible and
our theoretical prediction for the renormalization of the phonon velocity
agrees with experiment, the sound attenuation is very small and practically constant. This can easily be explained within our one-dimensional model.
Using Eqs.~(\ref{eq:Pi3}) and (\ref{eq:omegagamma}) we obtain
for the damping
 \begin{equation}
 \gamma_q= 
 \frac{\pi}{ 2 M \omega_q}
  \int_{ - \pi}^{\pi} \frac{d k}{2 \pi} 
 (f_k - f_{k+q}) 
| \Gamma_3 (  k , q ) |^2  
\delta ( \xi_k - \xi_{k+q} + \omega_q )  .
 \label{eq:damp}
 \end{equation} 
This expression is only finite if the absolute value of the maximal group velocity
$v_{\ast} = Z J b  $ of the fermions
exceeds the phonon velocity $c$. Because in   Cs$_2$CuCl$_4$
this condition is never satisfied ($c/(Jb)\approx6.8$), the attenuation of the $c_{22}$ mode vanishes
in our approximation. Higher orders in perturbation theory will give a finite
result, but it will involve more than two derivatives of the exchange coupling 
which are expected to be small.

On the other hand,  the condition $v_{\ast} > c$ can possibly be realized in some
other quasi-one dimensional quantum antiferromagnet. Let us therefore evaluate
Eq.~(\ref{eq:damp}) in the regime $v_{\ast} > c$. 
\begin{figure}[t]
\includegraphics[width=70mm]{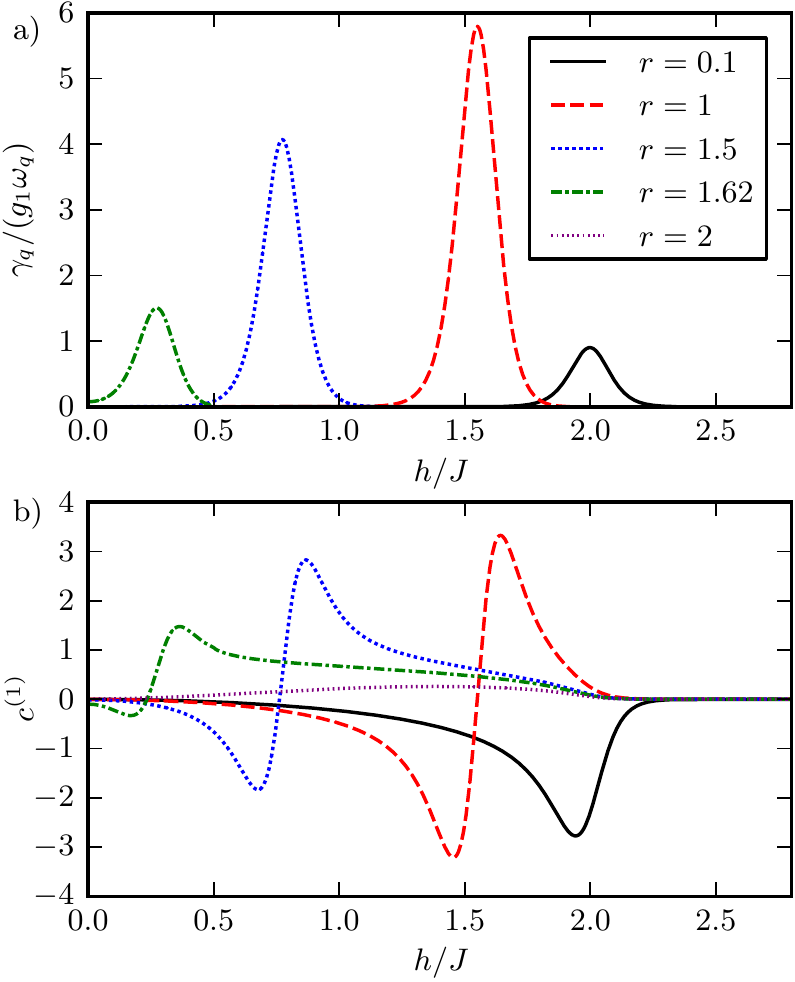}
  \caption{%
(Color online) a) Damping $\gamma_q$ for different values of $r=c/(Jb)$ as a function of the magnetic field at temperature $T/J=0.05$. b) Corresponding contribution $c^{(1)}$ to $\Delta c/c$ defined in Eq. (\ref{eq:deltac}).
}
  \label{fig:damp}
\end{figure}
In the long-wavelength limit $q \rightarrow 0$ we obtain
 \begin{equation}
  \frac{\gamma_q}{\omega_q } \sim
 \frac{ g_1 c J \Theta ( v^2_{\ast} - c^2 )}{2 v_{\ast}\sqrt{ 1 - c^2/v_{\ast}^2 }} 
 \left[  - f^{\prime} ( \xi_+ )  V_+^2    - f^{\prime} ( \xi_- )  V_-^2
 \right],
 \end{equation}
where $\xi_{\pm} =   J V_{\pm}-h$,
and $V_{\pm}  =   2s \pm Z  
 \sqrt{ 1 - c^2/ v_{\ast}^2 }.$
A numerical evaluation of this expression
is shown in the upper panel of Fig.~\ref{fig:damp}.

The damping exhibits strong peaks as  function of the magnetic field, corresponding
to the resonance conditions $\xi_{\pm } =0$ imposed by the
broadened delta-functions  $- f^{\prime} ( \xi_\pm )$ 
in Eq.~(\ref{eq:damp}). In the lower panel of Fig.~\ref{fig:damp}
we show that close to the resonance the corresponding shift $c^{(1)}$ in the phonon velocities
can exhibit a sign change depending on the value of $r=c/(Jb)$.

In summary, we have developed
a simple microscopic theory which explains ultrasound experiments
probing the propagation and the attenuation of the $c_{22}$ mode
in  the spin-liquid phase of Cs$_2$CuCl$_4$. 
Our basic assumption is that in the spin-liquid phase the elementary excitations are
one-dimensional fermions.
The excellent agreement between theory and experiments
shown in Fig.~\ref{fig:fitc} gives further support to the dimensional reduction
scenario advanced by Balents.\cite{Balents10}
It would be interesting to test our theoretical predictions for the ultrasound attenuation shown in Fig.~\ref{fig:damp} using suitable antiferromagnetic spin chains with sufficiently small phonon velocities.

We thank Andreas Kreisel for detailed comments on the manuscript. Financial support by the DFG via
SFB/TRR49 is gratefully acknowledged.


\begin{thebibliography}{99}
\bibitem{footnote1}
The phonon velocity $c$ in $x$ direction should not be confused with the lattice parameter $c$ in Fig. \ref{fig:lattice}. From here on, $c$ denotes the phonon velocity.
\end{thebibliography}
\end{document}